\def\vday{{3/13/98}}
\documentstyle[12pt]{article}
\def\baselinestretch{1.5}
\begin{document}
\thispagestyle{empty}
\null\vskip -1cm
\centerline{
\vbox{
\hbox{March 1998}\vskip -9pt
\hbox{\em\bf\vday}\vskip -9pt
\hbox{hep-ph/9803273}\vskip -9pt
     }
\hfill
\vbox{
\hbox{UICHEP-TH/98-2}
     }     } \vskip 1cm

\centerline
{\large \bf
A Simple Charged Higgs Model of Soft CP Violation }
\centerline
{\large \bf without Flavor Changing Neutral Currents
}
\vspace{1cm}
\centerline           {
David Bowser-Chao$^{(1)}$,
Darwin Chang$^{(2,3)}$,
and
Wai-Yee Keung$^{(1)}$ }
\begin{center}
\it
$^{(1)}$Physics Department, University of Illinois at Chicago,
IL 60607-7059, USA\\
$^{(2)}$Physics Department,
National Tsing-Hua University, Hsinchu 30043, Taiwan, R.O.C.\\
$^{(3)}$Institute of Physics, Academia Sinica, Taipei, R.O.C.\\
\vspace{.5cm}
\end{center}

\begin{abstract}
We propose a model of soft CP violation in which the  CP violating
mechanism naturally lies only in the charged Higgs sector.  The charged
Higgs mechanism not only accounts for the measured value of the
CP-violating parameter $\epsilon$ but also accommodates the current
limits on $\epsilon'/\epsilon$.  Our model naturally prevents tree-level
Flavor-Changing Neutral Currents (FCNCs) of any kind.  Unlike the
Weinberg-Branco Three-Higgs Doublet Model, the deviation from the
Standard Model rate for $b\to s\gamma$ is small.  Furthermore, leading
contributions to the electron (neutron)  electric dipole moment are
non-zero beginning at the three (two) loop level.  Surprisingly similar
to the Standard Kobayashi-Maskawa Model, our model is of milliweak
character but with seemingly superweak phenomenology.
\end{abstract}
\vspace{1in}
\centerline{Submitted to  {\it Physical Review Letters}}
\vspace{1in}
\centerline{PACS numbers: 11.30.Er, 14.80.Er \hfill}
\newpage

\section*{Introduction}

%------------+-----------------------------------------------------------
%   I. Introduction
%-----------------------------------------------------------------------

\quad
\vskip -1cm

Three decades after its surprising discovery in the kaon
system\cite{ccft}, CP violation has remained mysterious.  A desire for
deeper insight into its origin is the driving force behind many ongoing
experiments and even the construction of new machines such as the two B
Factories. While a profound understanding may yet be lacking, several
mechanisms have been suggested to explain   observed CP violation (i.e.,
$\epsilon \ne 0$) within a gauge field theory. Kobayashi and
Maskawa(KM)\cite{km} proposed a third generation of fermions, so that CP
violation would arise from the mixing of the three quark generations
and is manifested by a single phase in the Cabibbo-Kobayashi-Maskawa
(CKM) quark mixing matrix.
Since then, many other mechanisms have been put forth, including new gauge
interactions\cite{gauge},  neutral Higgs exchange\cite{neutral},
supersymmetric partners\cite{susy}, and  charged Higgs
exchange\cite{weinberg,branco}.
However, the KM model has the distinguishing feature
that its mechanism is of milliweak strength, though its
phenomenology is manifestly superweak\cite{superweak}, consistent with
current CP related data. Such intricate character has also been the driving
force behind the desire to find non-superweak CP violation in the $B$ systems.

The leading model for the charged Higgs mechanism of CP violation has
long been the Weinberg Three-Doublet Model of CP
violation\cite{weinberg}, which became even more intriguing after
Branco\cite{branco} proposed a version in which CP violation is softly
or spontaneously broken. This scheme naturally avoids tree-level flavor
changing neutral currents.  Without hard CP violation the CKM matrix is
purely real (the KM mechanism is inoperative); CP violation in the kaon
system instead results from charged Higgs exchange.  Many   weaknesses
of the Weinberg-Branco Model, however, have since been identified. Sanda
and Deshpande pointed out\cite{sanda}   that short distance
contributions to $\epsilon$, if  dominant, would lead to a larger
$\epsilon'/\epsilon$ than experimentally allowed, although it was
subsequently demonstrated that long distance contributions to $\epsilon$
could be large enough to avoid this difficulty\cite{chang}. More
recently, however, it has become clear that this model has other
problems. A charged Higgs light enough to account for the observed
$\epsilon$ has already been excluded by the LEP
experiments\cite{edm}. The large  neutron electric dipole
moment\cite{edm} (EDM)  and  substantial rate for $b \rightarrow s
\gamma$\cite{bsg} predicted are also contradicted by data,
leading several authors\cite{edm,bsg} to rule out this model.

As an illustrative model for  charged Higgs   CP violation, the
Weinberg-Branco Model also has the shortcoming that its neutral Higgs
sector naturally also contains CP violation, which is usually ignored in
the literature to simplify analysis and highlight the charged Higgs
mechanism. However, for flavor conserving CP odd observables (e.g., the
neutron EDM), the neutral Higgs contribution generically can be
competitive with that from charged Higgs exchange.

In this letter we propose an alternative model that may serve as a
generic example in which the  charged Higgs mechanism of CP violation
naturally dominates completely over other mechanisms. CP is broken
softly  or spontaneously so that the KM mechanism is inoperative.
Tree-level flavor changing neutral currents are automatically absent,
and the neutral Higgs sector is CP conserving at tree level. As in the
KM Model, the quark and electron EDMs
 are severely suppressed.  The
electron EDM vanishes at the two-loop level, while the first non-zero
contribution to the quark EDMs is at two loops. In contrast to the
Weinberg-Branco model, our model easily satisfies other experimental CP
violation constraints as well as the rate  for $b\to s \gamma$. Finally,
the parameter $\theta_{\rm{QCD}}$ vanishes at tree-level, since we
disallow hard CP breaking; we shall see that radiative corrections are
mild and consistent with the limit on a non-zero $\theta_{\rm{QCD}}$.

For most of this letter, we shall assume that CP is broken  softly. One
can also modify our model  to break CP spontaneously by introducing at
least one additional CP odd scalar boson, as discussed toward the end of
this work, with the bulk of the phenomenology unchanged.

\section*{General Formalism}

\quad \vskip-1cm

The Weinberg-Branco Model augments the Standard Model (SM) with
additional Higgs $SU(2)_L$ doublets, which are responsible for kaon
system CP violation; in this model, then, since the charged Higgs sector
must break CP, so also must the neutral Higgs sector. To mandate charged
Higgs exchange as the dominant CP violation mechanism we instead
introduce only additional $SU(2)_L$ singlets of quarks and scalars to
the theory. The simplest model for our purposes requires two additional
charged Higgs singlets, $h_\alpha (\alpha=1,2)$ and a vectorial pair of
heavy quark fields, $Q_{L,R}$, of electromagnetic charge $-{4\over3}$.
This vector quark charge assignment avoids fractionally charged hadrons.
Relevant new terms in the Lagrangian are:
\begin{equation}
{\cal L}_{h_i} =
  \left[
  (g \lambda_{i\alpha} \bar Q_L d_{iR} h_\alpha
+  M_Q \bar Q_L Q_R) + \hbox{h.c.} \right]
- (m^2)_{\alpha\beta} {h_\alpha}^{\dag} h_\beta
  - \kappa_{\alpha\beta}
        (\phi^{\dag} \phi-|\langle\phi\rangle|^2) \,h_\alpha^{\dag}
                       h_\beta  \;\;\label{eq:lagrangian}
\end{equation}
where $\phi$ is the Standard Model Higgs doublet, and $i$ is summed over
the  down quark flavors ($i=d,s,b$). The vector quark has purely
vectorial coupling to the photon and $Z$ boson, with respective charges
$(Q_Q, -Q_Q \sin^2\theta_W)$, while the charged Higgs  couples  with
charges $(Q_h, -Q_h \sin^2\theta_W)$  and $Q_Q = Q_d + Q_h$. The neutral
Higgs sector is identical to that in the Standard Model, with neither
flavor changing   couplings nor CP violation. The matrices $m^2$ and
$\kappa$ are hermitian.
Except for the discussion at the end, we assume that CP is broken softly
in this Lagrangian, implying a special basis where all the Yukawa
($\lambda, \kappa$) and the SM couplings are  real. We also require (see
below) that dim-3 couplings, namely $M_Q$, are also real. This leaves, as
in the KM model, only a single CP violating parameter:
Im$(m^2)_{12}$. We can diagonalize $(m^2)_{\alpha\beta}$ by a unitary
matrix $U_{\alpha i}$ which in general is complex:  $h_\alpha =
U_{\alpha i} H_i$, with $H_i$ the mass eigenstates. The quark-Higgs
interaction in the mass eigenstate basis is
\begin{equation}
{\cal L}_{QqH}=g\sum_{q=d,s,b}\xi_{qj}
	(\bar Q_L q_R)H^-_j \ +\ \hbox{h.c.}
\ ,
\label{eq:QqH}
\end{equation}
with $\xi_{qj} \equiv \lambda_{q\alpha} U_{\alpha j}$. The CP-violating
transit propagators\cite{weinberg} can be expressed as
$
\langle h_\alpha^{\dag} h_\beta \rangle
= \sum_{i,j=1,2} U_{i \alpha }^{\dag} U_{\beta j}
	\langle H_i^{\dag} H_j \rangle
= \sum_{i=1,2} U_{\beta i} U_{i \alpha }^{\dag}
	\langle H_i^{\dag} H_i \rangle
.
$
With $m_1$ $(m_2)$ the mass of the lighter (heavier) charged Higgs,
CP violation explicitly vanishes if  $m_1=m_2$. In the limit that
$m_2 \gg m_1$, these expressions reduce to $\langle h_\alpha^{\dag}
h_\beta \rangle = U_{\beta 1} U_{1 \alpha }^{\dag}
/(p^2-m_1^2+i\epsilon)\, , $ where $p$ is the momentum flowing in the
propagator. The rephasing-invariant measures of CP violation are then
${\cal A}_{qq'} = \lambda_{q\alpha}\lambda_{q'\beta}U_{\beta 1}
U_{\alpha 1}^{*} = \xi_{q1}^{*} \xi_{q'1}$ with $(q,q'=d,s,b)$,
 and
${\cal B} = \kappa_{\alpha \beta}U_{\beta 1} U_{\alpha 1}^{*}$.
For flavor changing processes, ${\cal A}_{qq'}$ plays the main role,
with ${\cal B}$ its counterpart in flavor conserving processes.

Before continuing, we   comment on the strong CP-violation parameter
$\theta_{\rm QCD}$. With CP symmetry imposed only on the hard (dim-4)
terms, the $\theta_{\rm QCD}$ parameter is naively zero at tree level,
but  $M_Q$ may still be complex. If so, alignment of the QCD vacuum
with this complex quark mass will generate a non-zero tree
level $\theta_{\rm QCD}$. To avoid this contribution, we simply impose
CP symmetry on both dim-4 and dim-3 terms. $M_Q$ will then be
real in the same basis that the tree-level $\theta_{\rm QCD}$
vanishes\cite{interplay}. A similar scheme can also be arranged if CP is
broken spontaneously (see below).  The first non-zero contribution to
$\theta_{\rm QCD}$ (occuring at two loops) will be discussed later.

\section*{Constraint from $\epsilon$}
With CP conservation modulo soft-breaking enforced, the CKM matrix is
real at tree level. Leading CP violating phenomena should be due solely
to the CP-violating phase in the charged Higgs sector. Making the usual
``$\pi \pi (I=0)$ dominance'' assumption, the CP violation parameter
$\epsilon$ is approximately
\begin{eqnarray}
\epsilon &   \simeq &\frac{e^{i\pi/4}}{\sqrt{2}}
\left(
	\frac{\hbox{Im}M_{12}}{2\hbox{Re}M_{12}}
	+\frac{\hbox{Im}A_0}{\hbox{Re}A_0}
\right)\ .
\end{eqnarray}
We shall postpone discussion of $A_0$, but will see later that in our
model, as in the KM Model,   the second term is negligible.
Experimentally,  $\epsilon \simeq 0.00226 \ \exp(i \pi/ 4)$. The $\Delta
S=2$ part of the effective Hamiltonian to one-loop (i.e., box diagrams)
can be written as:
\begin{equation}
\nonumber
{\cal H}^{\Delta S=2} =
\frac{G_F^2 m_W^2}{16\pi^2}
\sum_{I=R,L}C^I_{\Delta S=2}(\mu) O^I_{\Delta S=2}(\mu)
,\;\;\quad
O^{R,L}_{\Delta S=2} =\bar s \gamma_\mu (1\pm\gamma_5) d \,
\bar s \gamma^\mu (1\pm\gamma_5) d  \ .
\end{equation}
The $W$-boson diagrams yield a purely real Wilson coefficient
$C^L_{\Delta S=2}(\mu)$; CP violation in kaon matrix elements is due
solely to the operator $O^R_{\Delta S=2}$ rather than $O^L_{\Delta
S=2}$, in contrast to the KM model.  The complex coefficient
$C^R_{\Delta S=2}(\mu)$ is generated by the charged Higgs through a box
diagram with vertices given by Eq.(\ref{eq:QqH}).  At the scale $\mu=
M_Q$, we have
\begin{equation}
C^R_{\Delta S=2}(M_Q) =
  2 \xi_{d1} \xi_{s1}^* \xi_{d2} \xi_{s2}^{*}
	\frac{2m_W^2}{M_Q^2} \frac{f(x_2)-f(x_1)}{x_2-x_1}
+\sum_{i=1,2} (\xi_{di}\xi_{si}^{*})^2 \frac{2m_W^2}{M_Q^2}\,
{df \over dx}(x_i)
\ ,
\end{equation}
with $x_{1,2}=m_{H_{1,2}}^2/M_Q^2$,
$f(x)= (1-x+ x^2\log x )/(1-x)^2$, and $df/dx\-(1) = 1/3$.
Clearly, $C^R_{\Delta S=2}(M_Q)$ is real when $m_2 = m_1$ as it should
be.  For illustration, we shall take $m_2 \gg m_1$ and $m_1 = M_Q$, in
which case the first term is negligible and $C^R_{\Delta S=2}(M_Q) =
{2\over3}(\xi_{d1}\xi_{s1}^{*})^2m_W^2/M_Q^2$.

Following Ref.\cite{renormgroup} for the renormalization group evolution
and numerical evaluation of hadronic matrix elements to
leading order, we obtain
\begin{eqnarray}
C^R_{\Delta S=2}(\mu \le m_c) &=&
{\left[ \frac{\alpha_s(m_c)}{\alpha_s(\mu)} \right]}^{6/27}
{\left[ \frac{\alpha_s(m_b)}{\alpha_s(m_c)} \right]}^{6/25}
{\left[ \frac{\alpha_s(m_t)}{\alpha_s(m_b)} \right]}^{6/23}
{\left[ \frac{\alpha_s(M_Q)}{\alpha_s(m_t)} \right]}^{6/21}
C^R_{\Delta S=2}(M_Q) \nonumber \\
&\approx& 0.59 \,\alpha_s^{-2/9}(\mu)  \,C^R_{\Delta S=2}(M_Q) \ .
\end{eqnarray}
We will assume that the $W$-boson contributions dominate the real part
of all relevant matrix elements; analysis of $\epsilon^{'}/\epsilon$
below shows this to be consistent. We will thus take, {\it e.g}.,
$\hbox{Re}M_{12} = {1\over2}\Delta m_K$ from experiment, and have no
need of the explicit value of $W$-boson contributions to, e.g.,
$C^L_{\Delta S=2}$.
Let $M_{12}^R$ be the contribution of $O^R_{\Delta S=2}$ to the mass
matrix.  From the input parameters
$  B_K=0.75, F_K=160 \,{\hbox{MeV}}, m_K= 498\,{\hbox{MeV}} ,
    \Delta m_K = 3.51 \times  10^{-15} \hbox{GeV}   $
and the relation
$$
M_{12}^R =  \frac{1}{2m_K}
\langle \bar{K^0} | {\cal H}^{\Delta S=2} |K^0\rangle^{*}
= \frac{G_F^2 m_W^2}{16\pi^2} \frac{1}{2m_K} C_{\Delta S=2}^{R*}(\mu)
\langle \bar{K^0} | O^R_{\Delta S=2}(\mu) |K^0\rangle^{*}     \ ,
$$
\begin{equation}
\langle \bar{K^0} | O^R_{\Delta S=2}(\mu) |K^0\rangle =
	\hbox{${8\over3}$}\alpha_s(\mu)^{2/9} B_K  F_K^2 m_K^2    \ ,
\end{equation}
follows the numerical prediction
$ M_{12}^R / \Delta m_K = 1.2 \times 10^4 \  C_{\Delta S=2}^R(M_Q) $.
Demanding that the imaginary part of $M_{12}^R$ gives enough
contribution to $\epsilon$ and the corresponding real part gives just a
fraction $\cal F$ of the mass difference $\Delta m_K$ ({\it i.e.}
$ 2 \hbox{Re}( M_{12}^R ) = {\cal F} \Delta m_K$), we obtain
constraints on the Wilson coefficients:
$\hbox{Im}\ C_{\Delta S=2}^R(M_Q)= 2.7 \times 10^{-7} $ and
$\hbox{Re}\ C_{\Delta S=2}^R(M_Q)= 4.2 \times 10^{-5}{\cal F} $.
Again, with $m_2 \gg m_1, m_1=M_Q$, we then find
\begin{equation}
\hbox{Im} \left({\cal A}_{sd} / (0.049)^2 \right)^2 R_Q^2 =1      \ ,\quad
\hbox{Re} \left({\cal A}_{sd} / (0.049)^2 \right)^2 R_Q^2 =156 {\cal F}\ ;
\label{eq:dmineq}
\end{equation}
where $R_Q = 300 \hbox{ GeV}/M_Q$. The reasonable constraint $|{\cal F}|
< 1$ can be easily satisfied.

\section*{Constraints from $(\epsilon'/\epsilon)$ and $B^0$--$\bar {B^0}$
mixing}

The parameter $\epsilon'$ describes direct CP violation in the kaon
system. It is given in terms of the $2\pi$ decay amplitudes
$A_{0,2}= {\cal A}(K \to (\pi\pi)_{0,2})$, where the subscript indicates
the isospin of the  outgoing state. With
$\omega= |A_2/A_0|=0.045$,
$\xi=\hbox{Im}A_0/\hbox{Re}A_0$, $\Phi\approx \pi/4$, and
$\Omega=(1/\omega )\cdot (\hbox{Im}A_2/\hbox{Im}A_0)$,
\begin{equation}
 \epsilon' = -\frac{\omega}{\sqrt 2} \xi (1-\Omega) \exp(i\Phi) \ .
\end{equation}

The dominant contributions should be the gluon and electroweak penguins
mediated by  $H$ and  $Q$.  In contrast to the KM Model, the vector
coupling  of the vector quark means that the $Z$ boson penguin will be
suppressed by $O(m_K^2/m_Z^2)$ due to vector current conservation. The
gluon penguin contributes only to $A_0$, but the isospin-breaking
electromagnetic penguin (EMP) gives rise to both $A_0$ and $A_2$. Due to
its suppression  by $O(\alpha/\alpha_s)$, the latter affects $\epsilon'$
solely through its contribution to $\Omega$. Including the effects of
evolution from the vector quark mass down to the charm mass scale, we
estimate the EMP contribution to be $\Omega_{\rm EMP}
\stackrel{<}{\sim} O(1)$. There is an additional contribution to
$\Omega$ from $\eta, \eta'$ isospin-breaking, with $\Omega_{\eta-\eta'}
= 0.25$.  We shall ignore the electromagnetic penguin contribution here
(inclusion of the electromagnetic penguin will be studied
elsewhere\cite{UsAgain}), and set $\Omega=\Omega_{\eta-\eta'}$ to simply
the analysis.  The inclusion of $\Omega_{\rm EMP}$ will not change our
conclusion qualitatively.

The gluon penguin diagram, which involves the virtual vector quark $Q$
and the charge Higgs boson, produces an effective Hamiltonian at the
electroweak scale:
\begin{equation}
{\cal H}^{\Delta S=1}= (G_F/\sqrt{2}) \tilde{C}
( \bar s T^a \gamma_\mu(1+\gamma_5)d )
                       \times
                   \sum_q (\bar q T^a \gamma^\mu q )  \ ,
\end{equation}
\begin{equation}
\tilde{C}=\alpha_s
\sum_i{\xi_{di}\xi^*_{si}\over 6\pi}
       {m_W^2\over M_Q^2}F({m_{H_i}\over M_Q^2})
\ .
\end{equation}
\begin{equation}
F(x)=
      {x^2(2x-3)\log x\over  (1-x)^4}
  +   {16x^2-29x+7    \over 6(1-x)^3} ; \,   F(1)={3\over4}.
\end{equation}
Written in terms of the operators in Ref.\cite{renormgroup} (but of
flipped chirality),
$${\cal H}^{\Delta S=1}=
(G_F / \sqrt{2}) \sum_{i=3}^{6} \tilde{C}_i \tilde Q_i  \ ,  $$
with $\tilde{C}_{4,6} = \tilde{C}/4$,
 $\tilde{C}_{3,5} = -\tilde{C}/(4N_c) $,
   $Q_{3 (5)} = (\bar s_i d_i)_{V+A}
                \sum_q (\bar q_j q_j)_{V+A (V-A)}$
and $Q_{4 (6)} = (\bar s_i d_j)_{V+A}
                 \sum_q (\bar q_j q_i)_{V+A (V-A)}$,
where we have adopted the common no\-ta\-tion
$(\bar q q)_{V+A} (\bar s d)_{V-A}
= \bar q \gamma^\mu (1+\gamma_5) q \; \bar s \gamma_\mu (1 - \gamma_5) d$.

Again, for simplicity, we study the scenario that $m_2 \gg m_1$ and $m_1=M_Q
\simeq 300$ GeV.  Numerically,
$
\tilde{C}(\mu=300 \hbox{ GeV}) = 2.8\times 10^{-4}
(\xi_{di}\xi^*_{si})
R_Q^2
\ .
$
The Wilson coefficients are then run from $M_Q$ down to the charm mass
scale via the leading logarithm renormalization group
equations\cite{renormgroup}, so that $\tilde{C}_i(\mu=m_c) = r_i
\tilde{C}(\mu=300 \hbox{GeV})$, where $(r_3,\cdots, r_6) =
(-0.16,0.22,-0.036,0.51)$.
We note that  the two other  Wilson coefficients,
$\tilde{C}_1,\tilde{C}_2$, are not generated in the evolution.
Terms contributing to CP violation, and thus $\epsilon^{'}$,
in  $A_0$ are
$\langle (\pi\pi)_0 | {\cal H}^{\Delta S=1} |K \rangle =
(G_F/\sqrt{2}) \sum_{i=3}^{6} \tilde{C}_i \langle (\pi\pi)_0 |
\tilde Q_i |K \rangle$.
Using the
expressions for the matrix elements $\langle (\pi\pi)_0 | \tilde Q_i |K
\rangle$ found in Ref.\cite{renormgroup} at the scale $\mu=m_c= 1.3
\hbox{ GeV}$, we obtain
$$ \langle (\pi\pi)_0 | \{\tilde Q_3 \dots \tilde Q_6\}  |K \rangle
   (\mu=m_c) = \{0.012,0.19, -0.10, -0.30\} \hbox{ GeV}^{3} \ ,  $$
$$ \hbox{Im}A_0 = -
\hbox{Im}({\cal A}_{sd})
R_Q^2
\times 2.5 \times 10^{-10} \hbox{ GeV}  \ .
$$
For the weak couplings considered here, $\hbox{Re}A_0$ is approximated
quite well by the experimental value $|A_0| = 3.33\times 10^{-7} \hbox{
GeV}$, $\omega = 0.045$, so that
\begin{equation}
\left({\epsilon' \over \epsilon}\right)
= 1.9 \times 10^{-5} \
\hbox{Im}                 \left(
{{\cal A}_{sd} \over (0.049)^2}
\right)R_Q^2
= \pm 1.9 \times 10^{-5} \
\left(\sqrt{(156{\cal F})^2 +1} -156{\cal F}
\right)^{1/2} {R_Q \over \sqrt{2}} \ .
\end{equation}
The second equality is derived from constraints in Eq.(\ref{eq:dmineq}).
For $R_Q =1$  and ${\cal F} \approx 0$, $\epsilon'/\epsilon = 1.4 \times
10^{-5}$, which is somewhat smaller than, but certainly consistent with,
the results of the FNAL-E731 measurement of $(7.4 \pm 5.9) \times
10^{-4}$\cite{fnalE731}, but further from agreement with the CERN-NA31
result of $(23\pm 7)\times 10^{-4}$\cite{cernNA31}. If we relax the
constraint on the contribution to $\Delta m_K$ to allow ${\cal F}=-0.3$
(reflecting the uncertainty due to the large long-distance
contributions), then $\epsilon^{'}/\epsilon$ rises to $1.3 \times
10^{-4}$. If the omitted electromagnetic penguin contribution
$\Omega_{\rm {EMP}}$ turns out to be negative and important, it could
increase the predicted value of $\epsilon^{'}$ by perhaps as much as a
factor of two, still well below the experimental limit.

Another (much weaker) con\-straint to be con\-sidered is that from the
$B^0_{s,d}$ mass split\-ting\cite{renormgroup}. Proceeding in close
analogy to the calculation of the contribution to $\Delta m_K$, we
obtain:
\begin{equation}
\Delta M_{B^0} =
	2 \frac{G_F^2 m_W^2 }{16\pi^2} \frac{1}{2 m_B}\eta_B
	\left(\frac{2}{3}\;\hbox{Re}{\cal A}_{bd}^2\;
\frac{m_W^2}{M_Q^2}\right)
	\left( \frac{8}{3} B_B F_B^2 m_B^2 \right) \, ,
\end{equation}
where  again $m_2 \gg m_1$, $m_1=M_Q = 300$ GeV,  with the
renormalization group scaling factor $\eta_B=0.55$ evaluated as for
$\Delta m_K$, and $B_B=1$, $F_B=180$ MeV, $m_B= 5.28$ GeV. Given the
experimental value $\Delta M_{B^0} = 3.3\times 10^{-13}$ GeV, we have
\begin{equation}
\delta(\Delta M_{B^0}) /\Delta M_{B^0} = 1.1\times 10^{-3}\,
R_Q^2 \,
\hbox{Re}\left({\cal A}_{bd} /0.049^2\right)^2  \  .
\end{equation}
Even taking ${\cal A}_{bd}= (0.13)^2$, the fractional contribution
is only about $5\%$.

\section*{Other Constraints}
\noindent
{\it  $b \to s \gamma$}:
Because the operator  due to charged Higgs diagrams has  helicity opposite to
that generated in the Standard Model contribution, the two do not
interfere at amplitude level. Taking $m_2 \gg m_1=M_Q\simeq 300$ GeV:
\begin{equation}
{\delta B(b\rightarrow s\gamma)\over B(b\rightarrow s\gamma)_{\rm SM}}
=3.2\times 10^{-6}
\left|0.0389 \over {V_{tb}V^*_{ts}}\right|^2
R_Q^4
\left| \frac{ {\cal A}_{bs} }{ (0.049)^2} \
       \right|^2 \ .
\end{equation}
Furthermore, the relevant parameter ${\cal A}_{bd}$ is not subject
to constraints from $\epsilon$ or $\epsilon'$. If it is
of the same size as ${\cal A}_{sd}$, the deviation from the SM would
be negligible.

\noindent
{\it Strong CP and $\theta_{\rm QCD}$}:
There are no tree level complex quark masses  in our
model, and $\theta_{\rm QCD}$ is only induced starting at the two-loop
level, via generation of complex down-flavor quark masses. A typical
diagram is shown in Fig.~1; in contrast to $\epsilon$ and
$\epsilon^{'}$, this effect does not require more than one  flavor of
down-quark. Roughly, $\theta_{\rm QCD}\sim g^2 {\cal A}_{dd}\; {\rm
Im}{\cal B}/(16\pi^2)^2$. 
The present constraint, $\theta_{\rm QCD} < 10^{-9}$, 
can easily be accommodated, assuming a
moderately small  $\kappa$.

\noindent
{\it Neutron electric dipole moment}:
There is no one-loop diagram to produce the electric dipole moment
(EDM) of the light quarks, so    our model is very weakly
constrained  by neutron EDM limits. A down-flavor quark EDM,
however, is generated at the two-loop level, in parallel with
the generation of complex down quark masses discussed above.  A typical
contribution is given by Fig.~1, except with an external photon is
attached to internal charged lines. An estimate of the two loop
contribution is consistent with the current experimental bound.

\noindent
{\it Electron electric dipole moment}:
Unlike the down-flavor quarks, the electron couples only very indirectly
with the CP violating sector. The  electron EDM vanishes at the two loop
level. We expect the three-loop level contribution to be insignificantly
small.

\noindent
{\it Decay of new particles}:
In this model, $h$ and $Q$ can be assigned a new
conserved quantum number which guarantees a lightest exotic particle,
either $H_1$ or $Q$. A stable charged Higgs would lead to events with
possibly large missing transverse energy, while a stable vector quark
might be detected through formation of its  bound states\cite{tom}.
Alternatively, one can ignore this quantum number, so that an
additional interaction, $h_\alpha L_i L_j$, should be present,
which can lead to $H^-$ (on-shell or off-shell) decays into $l^-\nu$.
Even in this case, lepton number is still conserved, just as in the
Standard Model, since the vector
quark and charged Higgs will naturally carry the lepton number ($L=\pm
2)$.
Another way for $H$ to decay is to introduce a second
Higgs doublet and let $H$ couple  to two different Higgs doublets.  In
that case $H$ can decay into a neutral Higgs, plus a charged Higgs which in
turn decays into ordinary quarks and leptons.

\section*{Spontaneously Broken  CP symmetry}

We shall comment on the corresponding model in which CP is broken
spontaneously.  This can be implemented by adding a CP-odd scalar, $a$,
which develops a non-zero vacuum expectation value (VEV) and breaks CP.
However, this scalar will in general couple to $\bar{Q_L}Q_R$ and give
rise a complex tree level vector quark mass   and, therefore, a tree
level $\theta_{\rm QCD}$.  To avoid this, one can add another CP-even
scalar singlet, $s$, and impose  discrete symmetries which change the
signs of either or both $a$ and $s$ and nothing else.  As a result, a
term such as $ i a \bar Q \gamma_5 Q$ is forbidden and the only
additional term relevant for CP violation is $i\left[ s \,a
\,({h_1}^{\dag} h_2 - {h_2}^{\dag} h_1) \right]$.  This extra term will
give rise to complex $(m^2)_{12}$ after both $s$ and $a$ develop VEVs
and break CP. Note that before breaking CP spontaneously, there are two
possible definitions of CP symmetry, depending on which of $s$ and $a$
are defined to be CP odd; this is why both must develop develop VEVs in
order to break CP.  The extra neutral Higgs will  of course mix with the
SM Higgs, but since $a$ does not couple to fermions
directly, it will have scalar-pseudoscalar coupling to fermions only at
the loop level. As a result, its contribution to any CP violating
phenomenology will be small.

\section*{Conclusion}
We have proposed a model whose CP violation is solely mediated by
charged Higgs bosons.   The model is surprisingly similar to the KM
model in the sense that the CP-breaking mechanism is seemingly
milliweak, while its phenomenology (as studied here) is quite
superweak-like. The phenomenological distinction between the two  will
likely be made clear in  experiments planned for the B factory;  although
our model predicts a real CKM matrix, with corresponding 
collapse of the unitarity KM
triangle, new CP violating contributions will be contained in all the B
decay processes designed to measure this  triangle. A careful and
detailed analysis of such issues is clearly necessary and  is in
progress\cite{UsAgain}.

D.~B.-C. and W.-Y.~K. are supported by a grant from the Department of
Energy, and D.~C. by a grant from the National Science Council of
R.O.C.
We thank
M. Adams,
W. Bardeen,
G. Boyd,
P. Cho,
B. Grinstein,
J. Hughes,
T. Imbo,
R. Mohapatra,
and
L. Wolfenstein
  for useful discussions.
D.~C. also wishes to thank the High Energy Group and Physics Group of
Argonne National Laboratory its hospitality while this work was in progress.

\subsection*{Figures}

\begin{itemize}
\item[Fig. 1.]  A typical digram contributing to
a complex  down-flavor quark mass. A chiral rotation transforms this
contribution into the effective $\theta_{\rm QCD}$ term.
The cross represents the CP violating insertion
of $(m^2)_{12}$.
The same diagram, when attached with an external photon line, produces
an EDM for the the $d$-quark.

\end{itemize}

%\newpage

\def\baselinestretch{1.2}

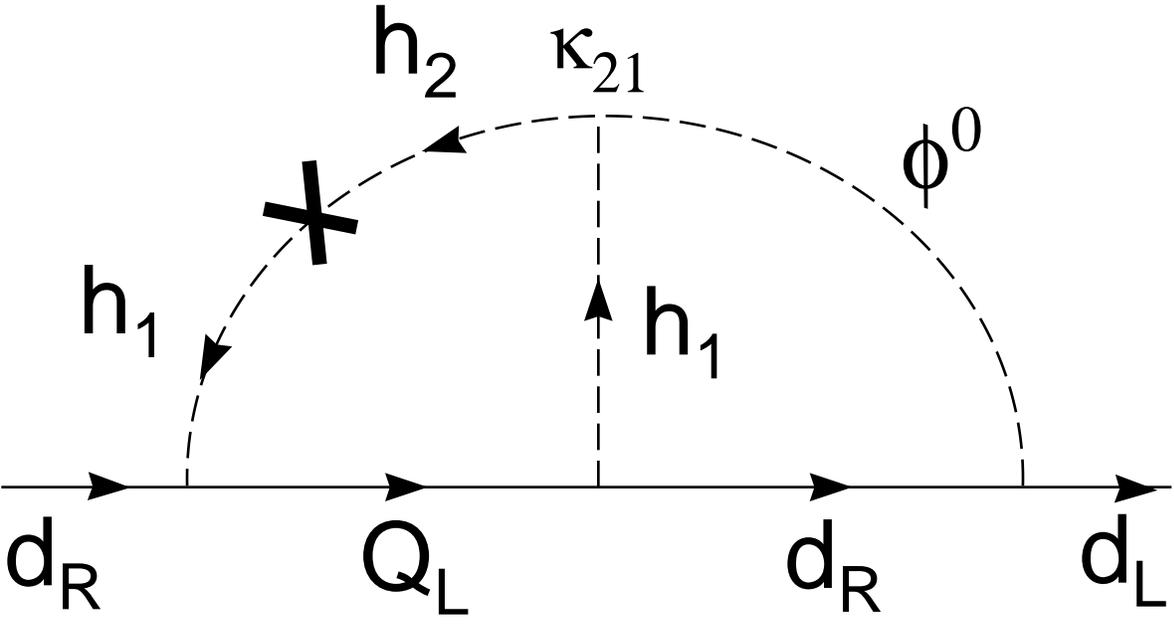
\end{document}